\documentclass[twocolumn,aps,prl,preprintnumbers,superscriptaddress]{revtex4}
\usepackage{}
\usepackage{bbm}
\usepackage[latin9]{inputenc}
\usepackage{amsmath}
\usepackage{amssymb}
\usepackage{graphicx}
\usepackage{mathrsfs}
\usepackage{amsfonts}
\usepackage{amsthm}
\usepackage{color}
\usepackage{txfonts}
\usepackage{lipsum}
\usepackage{gensymb}
\usepackage[colorlinks=true,citecolor=blue,linkcolor=blue,urlcolor=blue,anchorcolor=blue]{hyperref}%
\hypersetup{colorlinks=true,citecolor=blue,linkcolor=blue,urlcolor=blue}
\providecommand{\U}[1]{\protect\rule{.1in}{.1in}}
\makeatletter

\newcommand{\Rmnum}[1]{\expandafter\@slowromancap\romannumeral #1@}
\makeatother
\makeatletter
\@ifundefined{textcolor}{}
{
\definecolor{BLACK}{gray}{0}
\definecolor{WHITE}{gray}{1}
\definecolor{RED}{rgb}{1,0,0}
\definecolor{GREEN}{rgb}{0,1,0}
\definecolor{BLUE}{rgb}{0,0,1}
\definecolor{CYAN}{cmyk}{1,0,0,0}
\definecolor{MAGENTA}{cmyk}{0,1,0,0}
\definecolor{YELLOW}{cmyk}{0,0,1,0}
}
\makeatother
\begin{document}
\title{Type-II Weyl Excitation in Vortex Arrays}
\author{Z.-X. Li}
\affiliation{School of Electronic Science and Engineering and State Key Laboratory of Electronic Thin Films and Integrated Devices, University of Electronic Science and Technology of China, Chengdu 610054, China}
\author{X. S. Wang}
\affiliation{School of Physics and Electronics, Hunan University, Changsha 410082, China}
\author{Lingling Song}
\affiliation{School of Electronic Science and Engineering and State Key Laboratory of Electronic Thin Films and Integrated Devices, University of Electronic Science and Technology of China, Chengdu 610054, China}
\author{Yunshan Cao}
\affiliation{School of Electronic Science and Engineering and State Key Laboratory of Electronic Thin Films and Integrated Devices, University of Electronic Science and Technology of China, Chengdu 610054, China}
\author{Peng Yan}
\email[Corresponding author: ]{yan@uestc.edu.cn}
\affiliation{School of Electronic Science and Engineering and State Key Laboratory of Electronic Thin Films and Integrated Devices, University of Electronic Science and Technology of China, Chengdu 610054, China}
\begin{abstract}
Weyl-like magnon excitations in ordered magnets have attracted significant recent attention. Despite of the tantalizing physics and application prospects, the experimental observation of Weyl magnons is still challenging owing to their extraordinarily high frequency that is not accessible to the microstrip antenna technique. Here we predict gigahertz Weyl excitations in the collective dynamics of dipolar-coupled magnetic vortices arranged in a three-dimensional stacked honeycomb lattice. It is found that the inversion symmetry breaking leads to the emergence of the type-II Weyl semimetal (WSM) state with tilted dispersion. We derive the full phase diagram of the vortex arrays that support WSMs with both single and double pairs of Weyl nodes, and the topological insulator phase. We observe robust arc surface states in a dual-segment fashion due to the tilted nature of type-II WSMs. Our findings uncover the low-frequency WSM phase in magnetic texture based crystals that are indispensable for future Weyltronic applications.       
\end{abstract}

\maketitle
\emph{Introduction.}---The discovery of topological phase and phase transition has significantly deepened our understanding on the classification of matters \cite{KlitzingPRL1980,TsuiPRL1982,HasanRMP2010,QiRMP2011,ChiuRMP2016}. The exotic edge states are expected to have great potential applications in quantum computing and robust information processing \cite{LuNP2014,OzawaRMP2019,ZhangCP2018,YangPRL2015,HuberNP2016,MaNRP2019,LeeCP2018}. Depending on the band structure, there are two kinds of topological phases, one is the (higher-order) topological insulator (TI) \cite{BenalcazarS2017,SchindlerSA2018,XieNRP2021,LiPR2021} with gapped band structures, while the other one is the topological semimetal \cite{BurkovNM2016,ArmitageRMP2018,YanARCMP2017,GhorashiPRL2020,WangHXPRL2020,LuoNM2021,WeiNM2021} with gapless dispersion relations. Recently, Weyl semimetal (WSM) has received significant attention by the community because of its peculiar properties, such as the twofold linearly degenerated points (called Weyl points or nodes) between the conduction and valence bands, the arc surface state, and the chiral anomaly \cite{WanPRB2011,XuS2015,SoluyanovN2015,LuS2015,YangS2018,XiaoNP2015,RocklinPRL2016,ShiPRA2019,RafiNJP2020,DongPRR2021}. Due to the no-go theorem, the Weyl nodes (WNs) must come in pairs with opposite chiralities \cite{NielsenPLB1983,HosurCRP2013} and can be regarded as the source or sink of Berry curvature. Although WSMs have been intensively studied in electron system,   
their realization in magnetic system (especially by experiments) are relatively rare.

In magnetic ordered systems, the Weyl magnon  (quantized quasiparticle of spin wave) has been predicted theoretically in both ferromagnets and antiferromagnets with various crystal structures \cite{MookPRL2016,SuPRB2017_1,LiNC2016,JianPRB2018,SuPRB2017_2,OwerrePRB2018}. In spite of the theoretical progress, the experimental evidence for Weyl magnons is still lacking except for a very recent work \cite{ZhangPRR2020} claiming the observation of Weyl magnons by inelastic neutron scattering. The main reason that the observation of Weyl magnons is so difficult is owing to the high excitation frequency (from hundreds of gigahertz to several terahertz) which is not compatible with current microstrip antenna technique, let alone any practical applications. 
\begin{figure}
\begin{centering}
\includegraphics[width=0.5\textwidth]{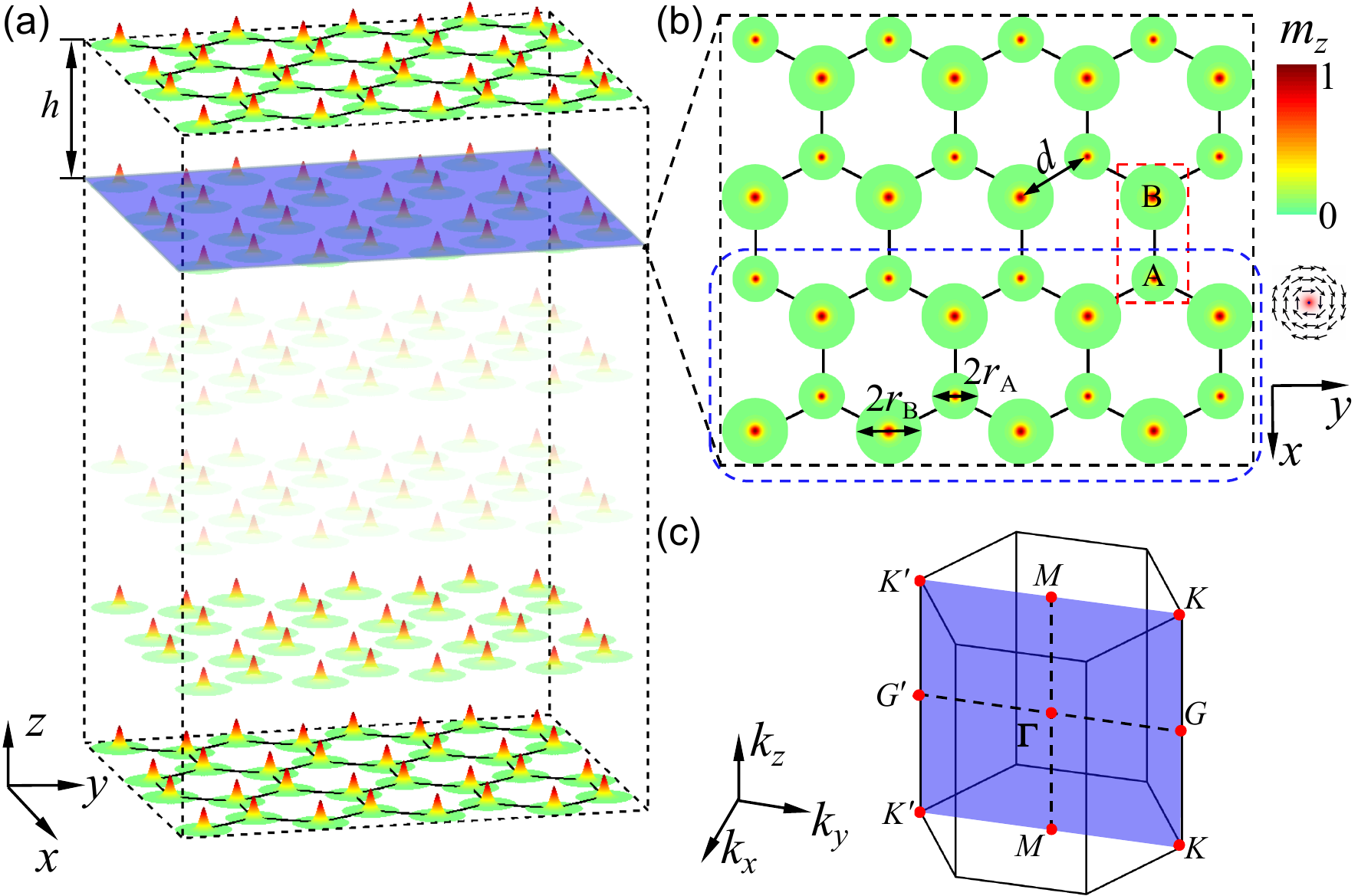}
\par\end{centering}
\caption{(a) Illustration of the 3D stacked honeycomb lattice composed of magnetic nanodisks of vortex states. (b) Zoomed in details of one layer. (c) The first Brillouin zone of the crystal.}
\label{Figure1}
\end{figure}Fortunately, in magnetic systems, magnetic solitons (such as vortex and skyrmion) carry much lower oscillation frequencies than magnons (by two or three orders of magnitude). It has been shown that the collective motion of magnetic solitons can exhibit the (higher-order) topological insulating phase \cite{LiPR2021,KimPRL2017,LiPRB2018,LiPRB2021_1,Linpj2019,LiPRA2020,LiPRB2020,LiPRB2021_2,LiAR2021,GoPRB2020}. However, the topological semimetal phase in magnetic soliton system is yet to be realized. Due to the low frequency (about 1 GHz) for vortex oscillations, one can expect a feasible experimental observation of vortex WSM phase by utilizing the microwave antenna, for instance. The realization of WSM phase in vortex lattice thus represents a key step for exploring the Weyl physics in pure magnetism and for designing three-dimensional (3D) Weyl spintronic devices.

In this paper, we theoretically study the collective dynamics of magnetic vortices in a 3D stacked honeycomb lattice (Fig. \ref{Figure1}). To realize the WSM phase, we break the sublattice symmetry of the system via introducing nanodisks with contrast geometry sizes and/or material parameters. By solving the coupled equations of motion for magnetic vortices, we obtain the phase diagram and predict the WSM phase of type-II with both one and two pairs of WNs. Furthermore, we demonstrate the vortex arcs formed by topologically nontrivial surface states, which is akin to the ``Fermi arcs'' for electrons.

\emph{Model.}---The collective dynamics of the 3D stacked honeycomb lattice of magnetic vortices [see Fig. \ref{Figure1}(a)] can be described by the Thiele's equation \cite{ThielePRL1973}:
\begin{equation}\label{Eq1}
\mathcal {G}\hat{z}\times \frac{d\textbf{U}_{j}}{dt}+\textbf{F}_{j}=0,
\end{equation}
where $\mathbf{U}_{j}= \mathbf R_{j} - \mathbf R_{j}^{0}$ is the vortex displacement $\mathbf{R}_{j}$ relative to its equilibrium position $\mathbf R_{j}^{0}$, $\mathcal {G}= -4\pi$$Qw M_{s}$/$\gamma$ is the gyromagnetic coefficient with $Q=\pm1/2$ the topological charge of the vortex, $w$ is the thickness of the nanodisk, $M_{s}$ is the saturation magnetization, $\gamma$ is the gyromagnetic ratio, and $\textbf{F}_{j}=-\partial {\mathcal{W}} / \partial \mathbf U_{j}$ is the restoring force with ${\mathcal{W}}$ being the total potential energy including the confining potential due to the nanodisk boundary and the coupling (including intralayer and interlayer) between nearest neighbor nanodisks: $\mathcal{W}=\sum_{j}\mathcal {K}_{j}\textbf{U}_{j}^{2}/2+\sum_{j\neq k}U_{jk}/2$ with $U_{jk}=\mathcal {I}_{\parallel}U_{j}^{\parallel}U_{k}^{\parallel}-\mathcal {I}_{\perp}U_{j}^{\perp}U_{k}^{\perp}+\mu_{j}\textbf{U}_{j}\cdot\textbf{U}_{k}$ \cite{ShibataPRB2003,ShibataPRB2004,GuslienkoAPL2005}. Here, $\mathcal {K}_{j}$ is the spring constant of $j$-th vortex, $\mathcal {I}_{\parallel}$ ($\mathcal {I}_{\perp}$) is the intralayer longitudinal (transverse) coupling constant, and $\mu_{j}$ is the interlayer coupling parameter. 

Imposing $\mathbf{U}_{j}=(u_{j},v_{j})$ and $\psi_{j}=u_{j}+iv_{j}$, Eq. \eqref{Eq1} can be re-written as: 
\begin{equation}\label{Eq2}
-i\dot{\psi}_{j}=\omega_{j}\psi_{j}+\zeta\sum_{k\in\langle j\rangle}\psi_{k}+\xi\sum_{k\in\langle j\rangle}e^{i2\theta_{jk}}\psi_{k}^{*}+\sum_{k\in\langle j'\rangle}\eta_{j}\psi_{k},
\end{equation}
where ${\omega}_{j}=\mathcal{K}_{j}/|\mathcal{G}|$, $\zeta=(\mathcal {I}_{\parallel}-\mathcal {I}_{\perp})/2\mathcal {|G|} $, and $\xi=(\mathcal {I}_{\parallel}+\mathcal {I}_{\perp})/2\mathcal {|G|}$; $\eta_{j}=\mu_{j}/|\mathcal{G}|$; $\theta_{jk}$ is the angle of the direction $\hat{\mathbf{e}}_{jk}$ from the $x$ axis with $\hat{\mathbf{e}}_{jk}=(\mathbf{R}_{k}^{0}-\mathbf{R}_{j}^{0})/|\mathbf{R}_{k}^{0}-\mathbf{R}_{j}^{0}|$; $\langle j\rangle$ and $\langle j'\rangle$ are the set of intralayer and interlayer nearest neighbors of $j$, respectively, with $j=A, B$ for different sublattices.

The WSM phase emerges only when at least one of the symmetries (time-reversal symmetry and inversion symmetry) is broken \cite{ArmitageRMP2018}. To obtain the WSM phase in our model, we break the sublattice symmetry. This can be achieved conveniently by setting contrast magnetic disk radius at different sublattices, i.e., $r_{A}\neq r_{B}$ [see Fig. \ref{Figure1}(b)]. We assume $\omega_{A}=\omega_{0}-\delta\omega$ and $\omega_{B}=\omega_{0}+\delta\omega$ with $\delta\omega$ being much smaller than $\omega_{0}$, which is justified when $|r_{A}-r_{B}|\ll(r_{A}+r_{B})/2$. Here $\omega_{A}$ and $\omega_{B}$ are the vortex gyration frequencies at A and B sublattices, respectively. By using the rotating wave approximation, Eq. \eqref{Eq2} can be recast as:
\begin{equation}\label{Eq3}
  \begin{aligned}
-i\dot{\psi}_{j}&=(\omega_{j}-\frac{3\xi^{2}}{2\omega_{0}})\psi_{j}+\zeta\sum_{k\in\langle j\rangle}\psi_{k}\\&-\frac{\xi^{2}}{2\omega_{0}}\sum_{s\in\langle\langle j\rangle\rangle}e^{i2\bar{\theta}_{js}}\psi_{s}+\sum_{k\in\langle j'\rangle}\eta_{j}\psi_{k},
  \end{aligned}
\end{equation}
where $\bar{\theta}_{js}=\theta_{jk}-\theta_{ks}$ is the relative angle from the bond $k\rightarrow s$ to the bond $j\rightarrow k$ with $k$ between $j$ and $s$, and $\langle\langle j\rangle\rangle$ represents the intralayer next-nearest neighbors of $j$.

For an infinite system, the unit cell can be chosen as the red dashed rectangle in Fig. \ref{Figure1}(b). Then the three basis vectors are $\mathbf{a}_{1}=\frac{3}{2}d\hat{x}+\frac{\sqrt{3}}{2}d\hat{y}$, $\mathbf{a}_{2}=\frac{3}{2}d\hat{x}-\frac{\sqrt{3}}{2}d\hat{y}$, and $\mathbf{a}_{3}=h\hat{z}$. Here parameter $d$ ($h$) denotes the distance between two intralayer (interlayer) nearest neighbor vortex cores. By considering the plane wave expansion of $\psi_{j}=\phi_{j}\text{exp}(i\omega t)\text{exp}[i(m\mathbf{k}\cdot\mathbf{a}_{1}+n\mathbf{k}\cdot\mathbf{a}_{2}+q\mathbf{k}\cdot\mathbf{a}_{3})]$, where $\mathbf{k}$ is the wave vector, and $m$, $n$ and $q$ are three integers, we obtain the system Hamiltonian:   
\begin{equation}\label{Eq4}
 \mathcal {H}=h_{0}\sigma_{0}+h_{x}\sigma_{x}+h_{y}\sigma_{y}+h_{z}\sigma_{z},
\end{equation}
with $\sigma_{0}$ the identity matrix, $\sigma_{x}$, $\sigma_{y}$ and $\sigma_{z}$ being the Pauli matrices. The parameters $h_{i}$ ($i=0,x,y,z$) can be explicitly expressed as:
\begin{equation}\label{Eq5}
\begin{aligned}
h_{0}&=\frac{\omega_{A}+\omega_{B}}{2}-\frac{3\xi^{2}}{2\omega_{0}}+\frac{\xi^{2}}{2\omega_{0}}[2\text{cos}(3k_{x}d/2)\text{cos}(\sqrt{3}k_{y}d/2)\\&+\text{cos}(\sqrt{3}k_{y}d)]+(\eta_{A}+\eta_{B})\text{cos}(k_{z}h),\\
h_{x}&=\zeta[1+2\text{cos}(3k_{x}d/2)\text{cos}(\sqrt{3}k_{y}d/2)],\\
h_{y}&=-2\zeta\text{sin}(3k_{x}d/2)\text{cos}(\sqrt{3}k_{y}d/2),\\
h_{z}&=\frac{\omega_{A}-\omega_{B}}{2}-\frac{\sqrt{3}\xi^{2}}{2\omega_{0}}[-2\text{cos}(3k_{x}d/2)\text{sin}(\sqrt{3}k_{y}d/2)\\&+\text{sin}(\sqrt{3}k_{y}d)]+(\eta_{A}-\eta_{B})\text{cos}(k_{z}h).
\end{aligned}
\end{equation}
Solving Eq. \eqref{Eq4} gives the eigenfrequencies of the system:
\begin{equation}\label{Eq6}
\omega_{\pm}=h_{0}\pm\sqrt{h_{x}^{2}+h_{y}^{2}+h_{z}^{2}}.
\end{equation}

\emph{Magnetic vortex WSMs.}---The band structure of the 3D vortex array is given by Eq. \eqref{Eq6}. To realize the WSM phase, we need to close the gap, i.e., $h_{x}=h_{y}=h_{z}=0$ at some specific points $\mathbf{k}$. It is straightforward to find four different points $\mathbf{k}_{1}^{\pm}=[0, 4\pi/3\sqrt{3}d, \pm\text{arccos}(g_{1})/h]$ and $\mathbf{k}_{2}^{\pm}=[0, -4\pi/3\sqrt{3}d, \pm\text{arccos}(g_{2})/h]$ in momentum space, which leads to the gapless band structure. Here $g_{i}=\frac{\omega_{B}-\omega_{A}}{2(\eta_{A}-\eta_{B})}+(-1)^{i}\frac{9\xi^{2}}{4\omega_{0}(\eta_{A}-\eta_{B})}$ and $i=1$, $2$. Near the points $\mathbf{k}_{1}^{\pm}$ and $\mathbf{k}_{2}^{\pm}$, by using the Taylor expansion, the effective Hamiltonian can be expressed as: 
\begin{equation}\label{Eq7}
\begin{aligned}
\mathbf{k}_{1}^{\pm}:  \mathcal{H}_{1}^{\pm}&=[\frac{\omega_{A}+\omega_{B}}{2}-\frac{9\xi^{2}}{4\omega_{0}}+(\eta_{A}+\eta_{B})g_{1}\\&\mp(\eta_{A}+\eta_{B})h\sqrt{1-g_{1}^{2}}p_{z}]\sigma_{0}-\frac{3}{2}\zeta dp_{y}\sigma_{x}\\&+\frac{3}{2}\zeta dp_{x}\sigma_{y}\mp(\eta_{A}-\eta_{B})h\sqrt{1-g_{1}^{2}}p_{z}\sigma_{z},\\
\mathbf{k}_{2}^{\pm}:  \mathcal{H}_{2}^{\pm}&=[\frac{\omega_{A}+\omega_{B}}{2}-\frac{9\xi^{2}}{4\omega_{0}}+(\eta_{A}+\eta_{B})g_{2}\\&\mp(\eta_{A}+\eta_{B})h\sqrt{1-g_{2}^{2}}p_{z}]\sigma_{0}+\frac{3}{2}\zeta dp_{y}\sigma_{x}\\&+\frac{3}{2}\zeta dp_{x}\sigma_{y}\mp(\eta_{A}-\eta_{B})h\sqrt{1-g_{2}^{2}}p_{z}\sigma_{z},
\end{aligned}
\end{equation}
\begin{figure*}[ptbh]
\begin{centering}
\includegraphics[width=0.82\textwidth]{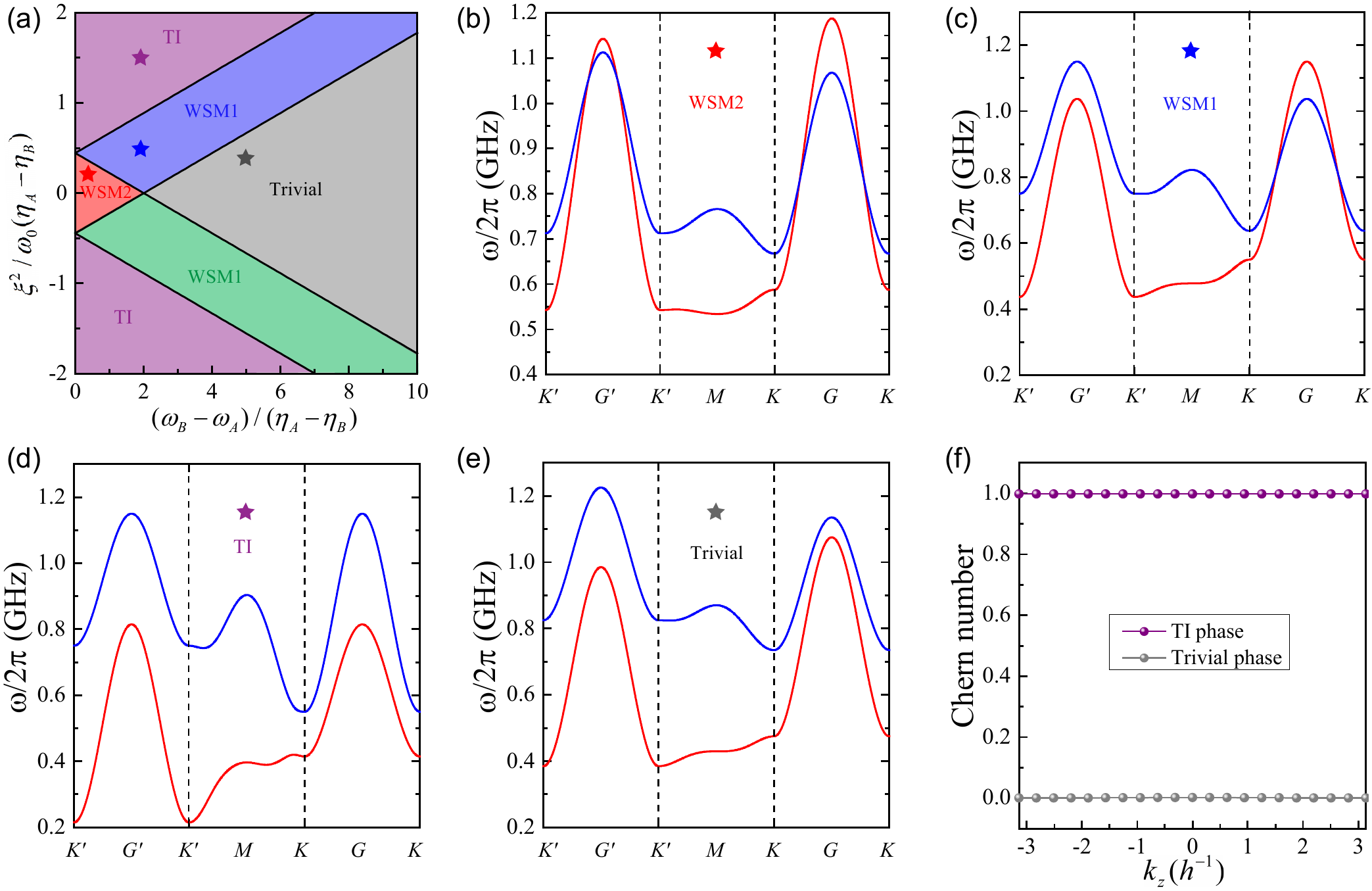}
\par\end{centering}
\caption{(a) Phase diagram of the system. Band structures along the path $K'-G'-K'-M-K-G-K$ in the momentum space for different parameters: (b) $(\omega_{B}-\omega_{A})/(\eta_{A}-\eta_{B})=0.5$, $\xi^{2}/\omega_{0}(\eta_{A}-\eta_{B})=0.2$, (c) $(\omega_{B}-\omega_{A})/(\eta_{A}-\eta_{B})=2$, $\xi^{2}/\omega_{0}(\eta_{A}-\eta_{B})=0.5$, (d) $(\omega_{B}-\omega_{A})/(\eta_{A}-\eta_{B})=2$, $\xi^{2}/\omega_{0}(\eta_{A}-\eta_{B})=1.5$, and (e) $(\omega_{B}-\omega_{A})/(\eta_{A}-\eta_{B})=5$, $\xi^{2}/\omega_{0}(\eta_{A}-\eta_{B})=0.4$. (f) Dependence of Chern number on $k_{z}$ for TI and trivial phases.}
\label{Figure2}
\end{figure*}where $\mathbf{p}=\mathbf{k}-\mathbf{k}_{i}^{\pm}$ ($i=1,2$). From Eqs. \eqref{Eq7}, one can clearly see that the conduction band and the valence band of the vortices oscillation are linearly touching at $\mathbf{k}_{1}$ (or $\mathbf{k}_{2}$), which is a typical feature of band crossing for WSMs. Furthermore, compared to the conventional $2\times 2$ Weyl Hamiltonian, our model contains the tilting term $(\eta_{A}+\eta_{B})\text{cos}(k_{z}h)\sigma_{0}$. According to the classification of WSMs \cite{LvRMP2021,WuPRB2017}, the ratio of the tilting term to the $z$-term $(\eta_{A}-\eta_{B})\text{cos}(k_{z}h)\sigma_{z}$ determines the type of WSM \cite{FootNote1}. Here the condition $|(\eta_{A}+\eta_{B})/(\eta_{A}-\eta_{B})|>1$ is always satisfied because coupling parameters $\eta_{A}>0$ and $\eta_{B}>0$. We therefore conclude that the semimetal phase reported in this work belongs to the type-II WSM.   

To guarantee the existence of WNs, one needs $|g_{1}|<1$ or $|g_{2}|<1$. In addition, we evaluate the topological Chern number ($\mathcal{C}$) for the gapped phase to derive the complete phase diagram of the system. It is worth noting that Chern number cannot be defined in 3D momentum space. Here we calculate the Chern number in $k_{x}-k_{y}$ plane for different $k_{z}$. If $\mathcal{C}=1$ ($\mathcal{C}=0$) for all $k_{z}$, the system is in topological (trivial) insulating phase. The phase diagram is divided into five regions, as shown in Fig. \ref{Figure2}(a). It can be clearly seen that there are three WSM phases: (i) The WSM in the blue region with one pair of WNs at $\mathbf{k}_{1}^{\pm}$ (WSM1); (ii) The WSM in the green region with one pair of WNs at $\mathbf{k}_{2}^{\pm}$ (WSM1); (iii) The WSM in the red region with two pairs of WNs at $\mathbf{k}_{1}^{\pm}$ and $\mathbf{k}_{2}^{\pm}$ simultaneously (WSM2). Figure \ref{Figure2}(f) plots the dependence of Chern number on $k_{z}$ for different insulating phases. It is indeed shown that Chern number is quantized to $1$ $(0)$ for parameters locating in the purple (gray) region.

To demonstrate the difference between various phases emerging in our system, we plot the bulk band structure along the path $K'-G'-K'-M-K-G-K$ in the first Brillouin zone [see Fig. \ref{Figure1}(c)] for different parameters \cite{FootNote2}, with results plotted in Figs. \ref{Figure2}(b)-\ref{Figure2}(e). It shows that when the system is in the WSM2 phase (red region), the two bands have the crossings at points $\mathbf{k}_{1}^{\pm}$ and $\mathbf{k}_{2}^{\pm}$ simultaneously [see Fig. \ref{Figure2}(b)]. While the bands only cross at $\mathbf{k}_{1}^{\pm}$ when the system is in the blue WSM1 phase [see Fig. \ref{Figure2}(c)]. Obviously, the bands will exhibit crossings at $\mathbf{k}_{2}^{\pm}$ when the system is in the green WSM1 phase (not shown). Besides, the two bands never touch each other in the TI and trivial phases [see Figs. \ref{Figure2}(d) and \ref{Figure2}(e)]. 

Next, we show that the Berry curvature contains pairs of monopoles at the WNs. To this end, we calculate the distribution of Berry curvature in momentum space, which can be expressed as \cite{ShindouPRB2013_1,ShindouPRB2013_2,WangPRL2020,WangJAP2021}:  
\begin{equation}\label{Eq8}
\Omega_{n}^{r}(\mathbf{k})=i\text{Tr}\Bigg{[}P_{n}
\epsilon_{rst}\Bigg(\frac{\partial P_{n}}{\partial k_{s}}\frac{\partial P_{n}}{\partial k_{t}}-\frac{\partial P_{n}}{\partial k_{t}}\frac{\partial P_{n}}{\partial k_{s}}\Bigg)\Bigg],
\end{equation}     
where $\epsilon_{rst}$ is the third-order Levi-Civita symbol and $P_{n}$ is the projection matrix $P_{n}(\mathbf{k})=\phi_{n}(\mathbf{k})\phi_{n}(\mathbf{k})^{\dag}$ with $\phi_{n}(\mathbf{k})$ being the normalized eigenstate of \eqref{Eq4} for $n$-th band. \begin{figure}[ptbh]
\begin{centering}
\includegraphics[width=0.48\textwidth]{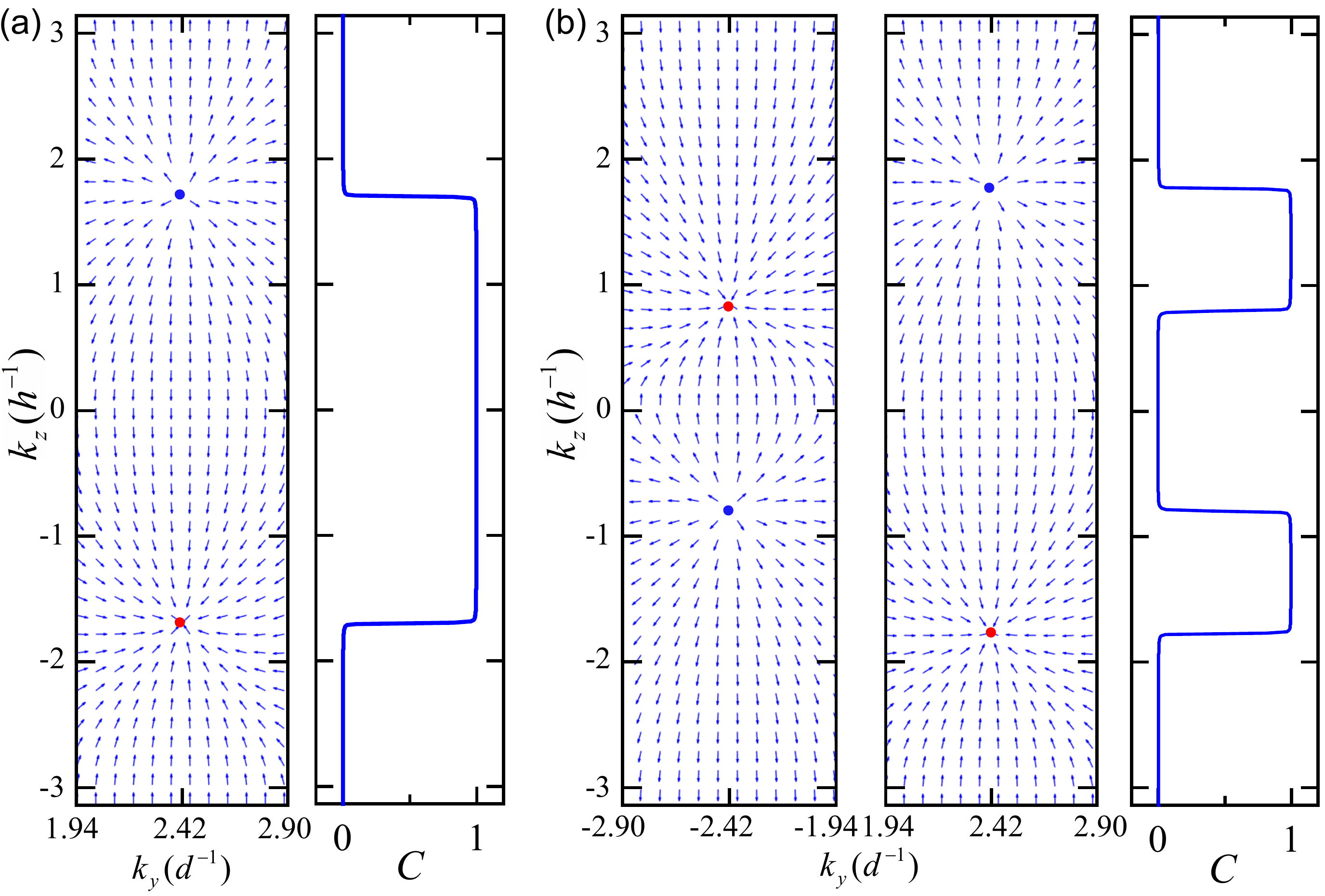}
\par\end{centering}
\caption{The Berry curvatures of the lowest band for WSM1 (a) and WSM2 (b) phases, respectively. The corresponding Chern number as a function of $k_{z}$ is shown in the right panels. }
\label{Figure3}
\end{figure}Here, for the sake of clarity, we project the Berry curvature onto the $k_{y}-k_{z}$ plane with $k_{x}=0$. The results are shown in the left panels of Figs. \ref{Figure3}(a) and \ref{Figure3}(b) (plotting the Berry curvature near the WNs only). The blue arrows represent the direction of Berry curvature. At the WNs, the Berry curvature indeed manifests the character of monopoles acting as sources or sinks. The Chern number as a function of $k_{z}$ are plotted in the right panels of Figs. \ref{Figure3}(a) and \ref{Figure3}(b). We find that the Chern number is quantized to 1 when $k_{z}$ is confined between the pair of WNs and to 0 otherwise, indicating that the topologically protected surface states appear only between the pair of WNs.

\begin{figure}[ptbh]
\begin{centering}
\includegraphics[width=0.48\textwidth]{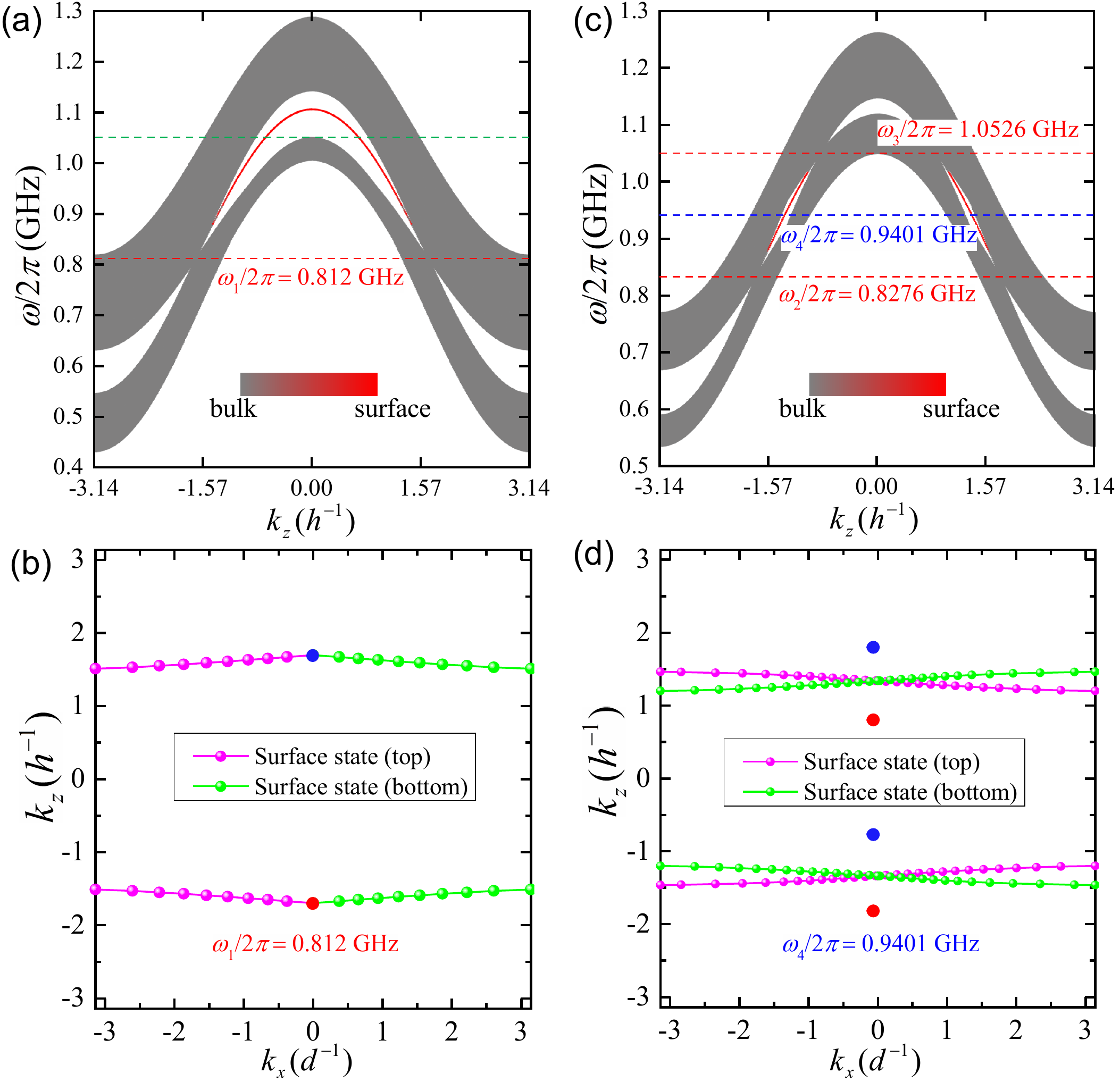}
\par\end{centering}
\caption{The band structures with the periodic boundary condition along $x$-axis and $z$-axis directions and the armchair termination along $y$-axis direction for the WSM1 (a) and WSM2 (c) phases. $\omega_{1}$, $\omega_{2}$, and $\omega_{3}$ are frequencies corresponding to the WNs and $\omega_{4}=(\omega_{2}+\omega_{3})/2$. (b) and (d) The corresponding isofrequency vortex arcs for the frequency $\omega=\omega_{1}$ and $\omega_{4}$, respectively.}
\label{Figure4}
\end{figure}

Although they are conducting in the bulk like normal metals, WSMs can support topologically protected surface states. To demonstrate this feature, we calculate the band structure of a finite system (armchair surface) with periodic boundary along $x$- and $z$-axes while open boundary along $y$-axis. The blue dashed box denotes the unit cell, as shown in Fig. \ref{Figure1}(b). Here we consider the unit cell containing 400 nanodisks. The corresponding band structure for WSM1 and WSM2 phases are shown in Figs. \ref{Figure4}(a) and \ref{Figure4}(c), respectively. The average position along $y$ direction of wave function $\langle{y}\rangle\equiv\sum_{j}R^{0}_{j,y}|\textbf{U}_{j}|^2/\sum_{j}|\textbf{U}_{j}|^2$ is also highlighted: the color closer to red indicates the mode more localized to the surface. Here $R^{0}_{j,y}$ is the equilibrium position of the vortices projected onto the $y$ axis. In what follows, we show that the in-gap modes are actually arc surface states.

\emph{Vortex arcs.}---To visualize the vortex arc, we plot the isofrequency curves in the $k_{x}-k_{z}$ plane with $\omega=\omega_{1}$ [for Fig. \ref{Figure4}(a)] and $\omega=(\omega_{2}+\omega_{3})/2$ [for Fig. \ref{Figure4}(c)] in Figs. \ref{Figure4}(b) and \ref{Figure4}(d), respectively. Here the magenta (green) lines represent the surface state confined at top (bottom) of the finite vortex crystal. Apparently, the surface states form arcs in a dual-segment style because of the tilted nature of type-II WSMs. The red and blue dots denote the WNs with opposite chiralities. It is worth mentioning that for WSM1 phase, if the frequency we chosen is not located at the WN, the two vortex arcs will gradually get close to each other with the increase of the frequency. When the frequency exceed a critical value [the green dotted line in Fig. \ref{Figure4}(a)], the two arcs touch each other. For WSM2 phase, however, no matter which frequency we choose between the WNs, there are always bulk states existing between the surface states. The two arcs can never touch each other.     

\emph{Discussion and conclusion.}---Both type-I and type-II WSMs can support exotic arcs, while their shapes are very different. For type-I WSMs, the surface states connect the pairs of WNs and form complete arcs for energy fixed at the WNs, because there is no bulk state between WNs. For type-II WSMs, however, the bulk states appear inevitably between the WNs because the Weyl cone is strongly tilted and Lorentz-violating. The surface states thus cannot connect the pairs of WNs and form partitioned arcs in a dual-segment fashion (see Fig. \ref{Figure4}).

In this work, we choose honeycomb vortex arrays to prove the emergence of the WSM phase in magnetic texture based metamaterials. It is straightforward that for other types of crystal structure (solitons), such as pyrochlore and kagome (magnetic bubbles and skyrmions), the novel WSM phases can emerge too. In addition, the polarity and chirality of all vortices are treated as identical in the present model. The effect from contrast polarities and chiralities on WSM state is an appealing research topic. The inversion symmetry along the $z$-axis is respected in this work. As this symmetry is broken, the vortex array may support higher-order WSM phase \cite{GhorashiPRL2020,WangHXPRL2020}, which is also an interesting issue for future study.

To conclude, we have studied the collective dynamics of magnetic vortices in 3D stacked honeycomb crystals. By breaking the sublattice symmetry, we predicted gigahertz Weyl semimetal and dual-segment arc surface states, the generation and detection of which are fully compatible with current microstrip antenna technique. Our work paves the way for realizing low-frequency Weyl excitations in ordered magnets that should be compelling for exploring fundamental Weyl physics and Weyltronic applications.
\begin{acknowledgments}
\emph{Acknowledgments.}---We thank C. Wang, Y. Su, Z. Wang, Z. Zhang, and H. Yang for helpful discussions. This work was supported by the National Natural Science Foundation of China (NSFC) (Grants No. 12074057, No. 11604041, and No. 11704060). Z.-X. Li acknowledges financial support from the NSFC (Grant No. 11904048) and the China Postdoctoral Science Foundation (Grant No. 2019M663461). X.S.W. acknowledges the support from the NSFC (Grant No. 11804045) and the Fundamental Research Funds for the Central Universities. 
\end{acknowledgments}


\begin{thebibliography}{99}
\bibitem{FootNote1}Without loss of generality, we write the tilt term as $H_\text{tilt}=b\text{cos}(k_{z})\sigma_{0}$, while the $z$-term as $H_{0}=t\text{cos}(k_{z})\sigma_{z}$, then the ratio of $b/t$ determines the type of WSMs. For $|b/t|<1$ ($|b/t>1|$), the WSM phase belongs to type-I (type-II). If $|b/t|=1$,  the corresponding semimetal phase becomes the type-III WSM.
\bibitem{FootNote2} Here we use the material parameters of permalloy (Py) with saturation magnetization $M_{S}=0.86\times10^{6}$ A m$^{-1}$ and exchange stiffness $A=1.3\times10^{-11}$ J m$^{-1}$. The thickness of the nanodisk $w=10$ nm, the gyration frequency $\omega_{0}=0.9\times2\pi$ GHz, interlayer coupling coefficient $\eta_{A}=0.15\times2\pi$ GHz, and $\eta_{B}=0.1\times2\pi$ GHz. For WSM1 phase [the blue star in Fig. \ref{Figure2}(a)], we choose $\omega_{A}=0.85\times2\pi$ GHz, $\omega_{B}=0.95\times2\pi$ GHz, ${I}_{\parallel}=3.8\times10^{-4}$ J m$^{2}$, and ${I}_{\perp}=2\times10^{-4}$ J m$^{2}$. For WSM2 phase [the red star in Fig. \ref{Figure2}(a)], we choose $\omega_{A}=0.8875\times2\pi$ GHz, $\omega_{B}=0.9125\times2\pi$ GHz, ${I}_{\parallel}=2.46\times10^{-4}$ J m$^{2}$, and ${I}_{\perp}=1.2\times10^{-4}$ J m$^{2}$. For TI phase [the purple star in Fig. \ref{Figure2}(a)], we choose $\omega_{A}=0.85\times2\pi$ GHz, $\omega_{B}=0.95\times2\pi$ GHz, ${I}_{\parallel}=6.5\times10^{-4}$ J m$^{2}$, and ${I}_{\perp}=3.5\times10^{-4}$ J m$^{2}$. For trivial phase [the gray star in Fig. \ref{Figure2}(a)], we choose $\omega_{A}=0.775\times2\pi$ GHz, $\omega_{B}=1.025\times2\pi$ GHz, ${I}_{\parallel}=3.45\times10^{-4}$ J m$^{2}$, and ${I}_{\perp}=1.73\times10^{-4}$ J m$^{2}$.
\end{thebibliography}
\end{document}